\documentclass{article}

\usepackage{amsmath}
\usepackage{authblk}
\usepackage{hyperref}
\usepackage{cite}
\usepackage{geometry}

\begin{document}

\title{Heun Functions and some of their applications in Physics  \footnote{This paper is a revised and many times updated version of the conference talk by the same author,  published in \textit{Proceedings of the 13th Regional Conference on
Mathematical Physics, Antalya, Turkey,27-31 October 2010}, edited
by U\u{g}ur Camc\i\ and Ibrahim Semiz, pp. 23-39, World Scientific (2013).} }
\author{M. Horta\c{c}su \footnote{E-mail: hortacsu@itu.edu.tr}}
\affil{Mimar Sinan Fine Arts University, \\ Department of Physics, \\
Istanbul, Turkey}
\maketitle
\begin{abstract}
\noindent Most of the theoretical physics known today is described by
using a small number of differential equations. For linear
systems, different forms of the hypergeometric or the confluent
hypergeometric equations often suffice to describe the system
studied. These equations have power series solutions with simple
relations between consecutive coefficients and/ or can be
represented in terms of simple integral transforms. If the problem
is nonlinear, one often uses one form of the Painlev\'{e}
equations. There are important examples, however, where one has to
use higher order equations. Heun equation is one of these
examples, which recently is often encountered in problems in
general relativity and astrophysics. Its special and confluent
forms take names as Mathieu, Lam\'{e} and Coulomb spheroidal
 equations. For these equations whenever a power series solution is
written, instead of a two way recursion relation between the
coefficients in the series, we find one between three or four
different ones. An integral transform solution using simpler
functions also is not obtainable.The use of this equation in physics and mathematical literature
exploded in the later years, more than doubling the number of papers with these solutions
in the last decade, compared to time period since this equation was introduced in 1889 up to 2008.
We use SCI data to conclude this statement, which is not precise, but in the correct ballpark.
 Here this equation will be
introduced and examples for its use, especially in general
relativity literature will be given.
\end{abstract}

\newpage

\section{Introduction}
\noindent Most of the theoretical physics known today is described by
using a small number of differential equations. If we study only
linear systems, different forms of the hypergeometric or the
confluent hypergeometric equations often suffice to describe the
system studied. These equations have power series solutions with
simple relations between consecutive coefficients and/ or can be
represented in terms of simple integral transforms. If the problem
is described in terms of  nonlinear differential equations, then
one often uses one form of the Painlev\'{e} equations.

\noindent There are important examples, however, where one has to
use higher order equations. Such an equation was  proposed by Karl
Heun in 1889 \cite {Heun}. This equation and its confluent forms
becomes indispensable in general relativity if one studies exact
solutions of  wave equations in the background of certain metrics.
A well known example is the Kerr metric \cite {Kerr}. Although it
is possible to solve the wave equations in the background of some
metrics in terms of hypergeometric functions or its confluent
forms, this is not possible for the much studied  Kerr metric. If
we also study even the trivially extended forms of some metrics by
adding a flat dimension to the existing metric, we may have to
solve the Heun equation to obtain the exact solution.

\noindent  Here we will find introduce the Heun equation and its
confluent forms and  mention some of the properties of the Heun
equation. Then we will give some examples in physics, mainly in
gravitational physics, where one can find many recent papers. This
part is  meant to be a survey of the work done  in the field of
\emph{General Relativity and Quantum Gravity} concentrating in the
last decades.  In another section we will give an example where
the Heun equation emerges from a trivial extension of a wave
equation in the background of the Eguchi-Hanson instanton metric
\cite {Eguchi}. We will end with some concluding remarks.

\section{Heun equation}
 Let us review some well known
facts about second order differential equations. Differential
equations are classified according to their singularity structure
\cite {Morse1,Ince}. If a differential equation has no
singularities over the full complex plane, it can only be a
constant. Singularities are classified as regular singular and
irregular singular points. If the coefficient of the first
derivative has at most single poles, and the coefficient of the
term without a derivative has at most double poles when the
coefficient of the second derivative is unity, this second order
differential equation has regular singularities, which gives us
one regular solution while expanding around this singular point.
In general the second solution has a  pole or a 
branch point singularity. If
the poles of these coefficients are higher, we have  irregular
singularities and the general solution has an essential
singularity \cite {Morse2}.

\noindent As stated in Morse and Feshbach \cite{Morse1} an example
of a second order differential equation with one regular singular
point is
\begin{equation}
\frac {d^{2}w}{dz^{2}} = 0 .
\end{equation}%
\noindent This equation has one solution which is constant. The
second solution blows up at infinity. The differential equation
\begin{equation}
\frac {d^{2}w}{dz^{2}} +k^{2} w= 0 .
\end{equation}%
has one irregular singularity at infinity which gives an essential
singularity at this point. The equation
\begin{equation}
z \frac {d^{2}w}{dz^{2}} + (1+a) \frac {dw}{dz}= 0 .
\end{equation}%
has two regular singular points, at zero and at infinity.

\noindent In physics an often used equation is the hypergeometric
equation
\begin{equation}
z(1-z)\frac {d^{2}w}{dz^{2}} + [c - (1+a+b)z] \frac{dw}{dz}-ab w=
0 .
\end{equation}%
This equation has three regular singular points, at zero, one and
infinity. Jacobi, Legendre, Gegenbauer, Tchebycheff equations are
special forms of this equation. When the singular points at z=1
and z equals infinity are "coalesced"  at infinity, we get the
confluent hypergeometric equation
\begin{equation}
z\frac {d^{2}w}{dz^{2}} + (c - z) \frac{dw}{dz} - a w = 0 .
\end{equation}%
with an essential singularity at infinity and a regular singularity
at zero. Bessel, Laguerre, Hermite equations can be reduced to
this form.

\noindent An important property of all these equations is that
they allow infinite series solutions about one of their regular
singular points where a recursion relation can be found between
two consecutive coefficients. This fact allows  one to have an
idea about the general properties of the solution, as the asymptotic
behaviour at distant points, the radius of convergence of the
series, etc.

\noindent A new equation was introduced in 1889 by Karl M. W. L.
Heun \cite{Heun}. This is an equation with four regular singular
points at zero, one, an arbitrary point f between zero and one and
infinity. This equation is discussed in the book edited by
Ronveaux \cite{Ronveaux}. Most of the general information we give
below is taken from this book. As discussed there, any equation
with four regular singular points can be transformed to the
equation given below:
\begin{equation}
\frac {d^{2}w}{dz^{2}} + [\frac {c}{z} + \frac{ d}{z-1} + \frac{
e}{z-f}] \frac{dw}{dz} - \frac{abz-q}{z(z-1)(z-f)}w = 0 .
\end{equation}%
There is a relation between the constants given as $ a+b+1 = c+d+e
$ .This relaion is not related to the regularity of the singularity at infinity.It just gives the exponents of the term
multiplying the series solution around infinity in terms of $u=1/z$
as $a,b$.

If we try to obtain a solution in terms of a power series, one can
not get a recursion relation between two consecutive coefficients.
We have a relation at least between three coefficients.

\noindent
 It is known that \cite{maier2}  any second order
differential equation with n regular singular points has a family
of $2^{n-1} n!$ local solutions, which splits into 2n sets of
$$2^{n-2} (n-1)!$$ equivalent expressions, each set defining one of
the two Frobenius solutions in the neighborhood of a singular
point. The n! factor comes from permuting the n singular points
and the $2^{n-1}$ factor from negating exponent differences. Maier
\cite{maier2} gave the list of 192 local solutions for the
Heun equation.

\noindent The set of transformations that can be applied to the a
Fuchian equation with n singular points to generate alternative
expressions for this equation has order $2^{n-1} n!$ and acts on
the parameter space of the equation. This group of transformations
is isomorphic to the Coxeter group $D_n$. These transformations
generate $2^{n-2} (n-1)!$ solutions. For the Heun case n=4, and
this group is isomorphic to $D_4$ , a group of order 192. These
transformations will be the combination of Mobius transformations
and transformations which multiply the desired solution by powers.

\noindent It turns out that the Mobius group PGL(2,C), which
takes x to $\frac {(Ax+B)} {(Cx+D)}$, for nonvanishing AD-BC, can
be used where $x$ takes values from the different singular points.
For Heun equation with four regular singular points, this
transformation takes each singular point to five other points,
which have zeroes at the same value. These points are given below:

\noindent $x, x/(x - 1), x/f, x/(x - f), (1 - f)x/(x - f), (f -
1)x/f(x - 1) ,$

\noindent $1 - x, (x - 1)/x,( x - 1)/(x - f), (x - 1)/(f - 1), d(x
- 1)/(x - f), f(x - 1)/(f - 1)x,$

\noindent $1/x, 1/(1 - x), f/x, f/(f - x), (f - 1)/(x - 1), (1 -
f)/(x - f),$

\noindent $(x - f)/x, (f - x)/a, (x - f)/(x - 1), (f - x)/(f - 1),
(x - f)/f(x - 1) , (f - x)/(f - 1)x$.

\noindent Any one of these transformations maps three of the four
points, 0,1,$f$, infinity, into 0,1, infinity, but generally
changes the value of $f$, which takes one of the six possible
values: $ f_1 = f, f_2 = 1-f, f_3 = 1/f, f_4 = 1/(1-f), f_5 =
f/(f-1), f_6=(f-1)/f $. Each value is taken four times.

\noindent Just recall the Heun equation:
\begin{equation}
\frac {d^{2}w}{dx^{2}} + [\frac {c}{x} + \frac{ d}{x-1} + \frac{
e}{x-f}] \frac{dw}{dx} - \frac{abx-q}{x(x-1)(x-f)}w = 0 ,
\end{equation}%
written in terms of the real variable $x$. One writes the solution
to the Heun equation in the form:
$$ y(x) = x^{r}
(x-1)^{s} (1-x/f)^{t} u(x).$$ This changes the form of the
differential equation. For
 (i) $ r=0$  or $1-c ,$
 (ii)$ s =0$ or $1-d,$
 (iii)
$t = 0$ or $1-e,$
 however, the resulting equation has the Heun form.
The values given above are the exponents at the singularities
\cite{Arscott},\cite{Ronveaux2}.

\noindent Of course, the parameters of the equations change. For
each such combination, say for $r=0$, there are four possible
values s and t can take, namely both equal to zero; $s=1-d, t=0;
s=0, t= 1-d; s=1-d, t= 1-e.$ Thus we get three more solutions for
each solution. Another factor of six comes from the six different
possible values $f$ can take. In total for expansions around a
single regular singular point, we have twenty four equivalent
solutions, obtained by simply transforming the original equation.

\noindent The presence of two different indices for expansion
around each singular point doubles the number of equivalent
solutions, resulting in 48 solutions for expansions around each
singular point. Four singular points multiplies this number by
four giving the total of 192 local solutions.

\noindent It turns out that for infinite set of values of the
parameter q, there are solutions which are analytic at 0 and at 1
. These are called \emph{Heun functions}, whereas those which are
analytic only at one point are called \emph{local Heun functions}
\cite{ronveaux11}.

\noindent For integer values of one of $a, c-a,d-a,e-a$,and for
special finite values of $q$, solutions analytic at three
singularities exist, the so called \emph{ Heun polynomials}. A
special case is for $a = -n, n=0,1,2$ and $q_{n,m} , m=0,1, ....,
n $, where $q_{n,m}$ are eigenvalues of a tridiagonal matrix, we
get the solution as a polynomial of degree n, which is analytic at
three singular points, 0,1 and $f$.
 \cite{ronveaux41}.

\noindent "No example has been given of a solution of Heun's
equation expressed in the form of a definite integral or contour
integral involving only functions which are , in some sense,
simpler" \cite{Ronveaux1}. This  statement  does not exclude the
possibility of having an infinite series of integrals  with
"simpler" integrands.

\noindent One can obtain different confluent forms of this
equation. When we "coalesce" two regular singular points, we get
the confluent Heun equation: The standard form of the confluent
form equation is given as \cite{Fiziev}.
\begin{equation}
{\frac{{d^{2}w}}{{dz^{2}}}}+\left( \alpha +{\frac{{\gamma+1}}{{z-1}}}+{\frac{{\beta+1}}{{z}}}%
\right) {\frac{{dw}}{{dz}}}+\left( {\frac{{\nu}}{{z-1}}}+{\frac{{\mu}}{{%
z}}}\right)w =0 .
\end{equation}
with solution
$$ HeunC (\alpha, \beta, \gamma, \delta,\eta,z).
$$
$$\delta = \mu+\nu-\alpha ( \frac{{ \beta+\gamma+2}}{{2}} ),$$
$$ \eta = \frac{{ \alpha(\beta+1) }}{{2}} - \mu - (\frac {{ \beta+\gamma+ \beta \gamma}}{{2}}). $$
Another version of this equation can be written as
\begin{equation}
\frac {d}{dz} (( z^2 -1)\frac {dw}{dz}) + [ -p^2 ( z^2-1) +2p\beta
z- \lambda - \frac{m^2+s^2+2msz}{(z^2-1)}]w = 0 .
\end{equation}%
Special forms of this equation are obtained in problems with two
Coulombic centers,
\begin{equation}
\frac {d}{dz} (( z^2 -1)\frac {dw}{dz}) + [ -p^2 ( z^2-1) +2p\beta
z- \lambda - \frac{m^2}{(z^2-1)}]w = 0 ,
\end{equation}%
whose special form, when $b=0$, is the spheroidal equation,
\begin{equation}
\frac {d}{dz} (( z^2 -1)\frac {dw}{dz}) + [ -p^2 ( z^2-1) -
\lambda - \frac{m^2}{(z^2-1)}]w = 0 .
\end{equation}%
Another form is the algebraic form of the Mathieu equation:
\begin{equation}
\frac {d}{dz} (( z^2 -1)\frac {dw}{dz}) + [ -p^2 ( z^2-1) -
\lambda - \frac{1}{4(z^2-1)}]w = 0 .
\end{equation}%
If we coalesce two regular singular points pairwise, we obtain the
double confluent form:
\begin{equation}
D^2 w + ( \alpha_{1} z + \frac{\alpha_{-1}}{z} ) Dw +[(B_{1} +
\frac{\alpha_{1}}{2} )z +( B_0 + \frac{\alpha_{1} \alpha_{-1}}{2}
) + ( B_{-1} - \frac{\alpha_{-1}}{2}) \frac{1}{z}) ] w = 0 .
\end{equation}%
Here $ D = z \frac{d}{dz} $.  We can reduce the new equation to
the Mathieu equation, an equation with two irregular singularities
at zero and at infinity if we reduce this equation to the form:
\begin{equation}
D^2 y + ( B z^2 + B_0 + B z^{-2} ) y = 0.
\end {equation}
 Another form
is the biconfluent form, where three regular singularities are
coalesced. The result is an equation with a regular singularity at
zero and an irregular singularity at infinity of higher order:
\begin{equation}
z^2 \frac {d^{2}w}{dz^{2}} + z \frac{dw}{dz}w +( A_0 + A_1 z + A_2
z^2 + A_3z^3 - z^4) w = 0 .
\end{equation}%
The anharmonic equation in three dimensions can be reduced to this
equation:
\begin{equation}
\frac {d^{2}w}{dz^{2}} +( E - \frac{\nu }{r^2} - \mu r^2 -\lambda
r^4 -\eta r^6 ) w = 0 .
\end{equation}%
In the triconfluent case, all regular singular points are
"coalesced" at infinity which gives the equation below:
\begin{equation}
\frac {d^{2}w}{dz^{2}} + (A_0 + A_1 z + A_2 z^2 -\frac{9}{4} z^4)
w = 0 .
\end{equation}%
These different forms are used in different  problems in physics.

\section{Some Examples of the Heun equation in Physical Applications}

\noindent In SCI we found about one hundred thirty  papers when
Heun functions were searched in the summer of 2010. Now, at the end of
 April  2018, the number exceeded 330. The number of
published articles in SCI more than doubled in the last eight
years.  More than three fourths of these papers were published in
the last ten years. The rest of the papers were published between
1990 and 2005, except a single paper in 1986 \cite{valent}.  These
numbers may differ depending on the institution where one uses the
SCI, since different universities in Turkey start their search
from different dates. We think we are still in the correct
\emph{ball park}. This shows that although the Heun equation was
found in 1889, it was largely neglected in the physics literature
until recently. Earlier papers on this topic are mostly articles
in mathematics journals. If one looks for books on this topic, published before the year 2000, one
finds out the the list of books is not very long. There is a book
edited by A.Ronveaux, which is a collection of papers presented in
the " Centennial Workshop on Heun's Equations: Theory and
Application. Sept.3-8 1989, Schloss Ringberg". It was published by
the Oxford University Press in 1995 by the title \emph{Heun's
Differential Equations} \cite {Ronveaux}. There are two books on
functions which are special cases of the Heun Equation:
\textit{Mathieusche Funktionen und Sphaeroidfunktionen mit
anwendungen auf physikalische und technische Probleme } by Joseph
Meixner and Friedrich Wilhelm Schaefke, published by Springer
Verlag in 1954 \cite { Meixner} and a Dover reprint of a book
first published in 1946, \textit{Theory and Applications of
Mathieu Functions} by N.W. McLachlan in 1963 \cite{McLachlan}.
Classical mathematical physics books, such as Morse and Feshbach
\cite {Morse1}, Whittaker and Watson\cite{Whittaker} or the
Batemann Manuscript \cite{Bateman},  have sections or
chapters on the special forms of the Heun equation like Mathieu,
Lam\'{e} or spheroidal functions. Some papers on different
mathematical properties of these functions can be found in
references \cite{schafke}- \cite{birkandant}.

\noindent A reason why more physicists are interested in the Heun
equation recently may be, perhaps, a demonstration of the fact
that we do not have simple problems in theoretical physics
anymore. Mathematical physicists have to tackle more difficult
problems, either with more difficult metrics or in higher
dimensions. Both of these extensions may necessitate the use of
the Heun functions among the solutions. We can give the
Eguchi-Hanson case as an example. The wave equation for the scalar
particle  in the background of the Eguchi-Hanson metric
\cite{Eguchi}  in four dimensions has hypergeometric functions as
solutions \cite{Nuri}, whereas the Nutku helicoid
\cite{Nutku,Petzold} metric, the next higher one, gives us Mathieu
functions \cite{Yavuz2}, a member of the Heun function set, if the
method of separation of variables is used to get a solution.  We
also find that the scalar particle, in the background of the Eguchi-Hanson metric, trivially extended to
five dimensions gives Heun type solutions. 
\cite{tolga2}.

\noindent Note that the problem does not need to be very
complicated to work with these equations. We encounter Mathieu
functions if we consider two dimensional problems with elliptical
shapes \cite{Morse3}. Let us use $ x= \frac{1}{2} a \cosh {\mu}
\cos{\theta}, y= \frac{1}{2} a \sinh {\mu} \sin{\theta}$, where
$a$ is the distance from the origin to the focal point. Then the
Helmholtz equation can be written as
\begin{equation}
{ \partial_{\mu \mu} \psi} + {
\partial_{\theta \theta} \psi} + \frac{1}{4} a^2 k^2 [
\cosh^2 {\mu} - \cos^2 {\theta} ] \psi=0
\end{equation}
which separates into two equations
\begin{equation}
\frac { d^2 H}{ d \theta^2} +( b - h^2 \cos^2{ \theta}) H =0,
\end{equation}
\begin{equation}
-\frac { d^2 M}{ d \mu^2} +( b - h^2 \cosh^2{ \mu}) M =0.
\end{equation}
The solutions to these two equations can be represented as Mathieu
and modified Mathieu functions.

\noindent If we combine different inverse powers of r, starting
from first up to the fourth, or if we combine the quadratic
potentials with inverse even powers of two, four and six, we see
that the solution of the Schrodinger equation involves Heun
functions \cite{bonorino4}. Solution to symmetric double Morse
potentials also needs these functions, like $ V ( x) = B^2 /4
sinh2 x - (s+ 1/2) B cosh x $ where $s= (0,1/2, 1,...)$
\cite{bonorino4}. Similar problems are treated in references
\cite{bonorino1}, \cite{bonorino2} and \cite{bonorino3}

o In atomic physics further problems such as separated double
wells, Stark effect, hydrogen molecule ion use these functions.
Physics problems which end up with these equations are given in
the book by S.Y. Slavyanov and S. Lay \cite{Slavyanov}. Here we
see that even the Stark effect, hydrogen atom in the presence of
an external electric field, gives rise to this equation. As
described in page 166 of Slavyanov's book, cited above ( original
reference is Epstein \cite{Epstein}, also treated by S.Yu
Slavyanov \cite{Slav1}),  when all the relevant constants, namely
Planck constant over $ 2 \pi $, electron mass and electron charge
are set to unity, the Schrodinger equation for the hydrogen atom
in a constant electric field of magnitude $F$ in the $ z $
direction is given by
\begin{equation}
\Big( \Delta + 2[ E-(F z - \frac {1}{r})] \Big) \Psi = 0 .
\end{equation}%
Here $\Delta$ is the laplacian operator. Using parabolic
coordinates, where the cartesian ones are given in terms of the
new coordinates by $ x= \sqrt { \xi \eta} cos{\phi}, y= \sqrt {
\xi \eta} sin {\phi} , z= \frac {\xi-\eta}{2} $ and writing the
wave function in the product form
\begin{equation}
\Psi = \frac{1}{\sqrt {\xi \eta}} V(\xi) U(\eta) exp ( im\phi),
\end{equation}%
we get two separated equations:
\begin{equation}
\frac{d^2 V}{d \xi^2 } +(\frac{E}{2} +\frac{\beta_1}{\xi} - \frac
{F}{4} \xi + \frac {1-m^2} {4\xi^2}) V(\xi) = 0,
\end{equation}%
\begin{equation}
\frac{d^2 U }{d \eta^2} + (\frac{E}{2} + \frac{\beta_2 }{\eta} +
\frac{F}{4} \eta +  \frac {1-m^2}{4\eta^2}) U(\eta) =0.
\end{equation}%
Here $\beta_1 $ and $\beta_2$ are separation constants that must
add to one. We note that these equations are of the biconfluent
Heun form.

\noindent The hydrogen molecule also is treated in reference
\cite{slav131}. When the hydrogen-molecule ion is studied in the
Born-Oppenheimer approximation, where the ratio of the electron
mass to the proton mass is very small, one gets two singly
confluent Heun equations if the prolate spheroidal coordinates $
\xi = \frac{ r_1 + r_2}{2c}, \eta= \frac{ r_1 + r_2}{2c} $ are
used. Here $c$ is the distance between the two centers. Assuming
\begin {equation}
\psi = = \sqrt {\xi \eta} V(\xi) U(\eta) exp ( im\phi ),
\end{equation}%
we get two confluent Heun equations:
\begin{equation}
\frac {d}{d\xi} \Big( (1-\xi^2) \frac {dV}{d\xi} \Big) + \Big(
\lambda^2 \xi^2 -\kappa \xi - \frac {m^2}{1-\xi^2}+ \mu \Big) V =
0 ,
\end{equation}
\begin{equation}
\frac {d}{d\eta} \Big( (1-\eta^2) \frac {dU}{d\eta} \Big) + \Big(
\lambda^2 \eta^2 - \frac {m^2}{1-\eta^2} + \mu \Big) U = 0 .
\end{equation}
\noindent  Some additional  physics papers with Heun type solutions include:

\noindent Three relatively recent papers which treat atoms in
magnetic fields:

o Exact low-lying states of two interacting equally charged
particles in a magnetic field are studied  by Truong and Bazzali
\cite{Truong1}

o The energy spectrum of a charged particle on a sphere under a
magnetic field and Coulomb force are studied by Ralko and Truong
\cite{Truong2}

o B.S. Kandemir presented an analytical analysis of the
two-dimensional Schrodinger equation for two interacting electrons
subjected to a homogeneous magnetic field and confined by a
two-dimensional external parabolic potential. Here a biconfluent
Heun (BHE) equation is used \cite{Kandemir}

o Arda and Sever, in one instance with Aydo\u{g}du studied
Schrodinger equation with different potentials and in two cases
found Heun and confluent Heun solutions \cite{Arda1, Arda2}.

In two papers Hammann et al \cite{Hammann1,Hammann2} solved the one-dimensional Schrodinger equation for position-dependent masses, and obtained  Heun solutions. The importance of these papers is the derivation and use of a relations between Heun functions which are functions of $z$ and $1-z$, which can be used for obtaining the reflection and transition amplitudes for scattering problems for waves  described in terms of Heun functions.

o Recently Ishkhanyan showed that the solution of the Schrodinger
equation for the $ V_0 / \sqrt{x} $ can be given as a derivative
of a triconfluent Heun function \cite{Artur}. In another paper, solution for the same potential  is given \cite{Artur2} as a linear combination of two confluent hypergeometric functions. For another potential which is an inverse square root near the origin and vanishes exponentially at infinity, solution is given in terms of linear combination of Gauss hypergeometric functions\cite{Artur3}. These potentials belong to the Heun class.

Downing showed that the solution to the one dimensional Schrodinger equation with a hyperbolic  double well potential is obtained by a transformation of the confluent Heun equation \cite{Downing}.

Hartmann and Portnoi calculated the bound modes of two-dimensional massless Dirac fermions confined within a hyperbolic secant potential \cite{Hartmann} 

Portnoi et al. continued studying the two-dimensional Dirac particles in two papers,  first confined in nonuniform magnetic fields, and second in Poschl-Teller waveguide\cite{Portnoi1, Portnoi2} in terms of confluent Heun functions.

o In a relatively recent work P. Dorey, 
\cite{Dorey} showed that equations in finite lattice systems also
reduce to Heun equations.

o Dislocation movement in crystalline materials, quantum diffusion
of kinks along dislocations are some solid state applications of
this equation. The book by S.Y. Slavyanov and S. Lay
\cite{Slavyanov} is a general reference on problems solved before
2000.

\noindent

We also cite a  recent mathematical application by A.M.
Ishkhanyan et al. where " total fifteen potentials for which the
stationary Klein-Gordon equation is solvable in terms of the
confluent Heun functions are presented.. Only nine of the
potentials are independent due to the  transposition symmetry of
regular singular points of  the equation. Four of these equations
can be reduced to the hypergeometric form. The remaining five
independent Heun potentials are four-parametric and have solutions
only in terms of irreducible confluent Heun functions \cite
{Artur13}. Prof Ishkhanyan expands the Heun solution in terms of hypergeometric functions
 and shows that the sum has only finite number of terms in his cases.
Prof. Ishkhanyan wrote additional papers after this
 one using the same method for other potentials. We will not comment on them,
 however, since from this point on, we will confine ourselves only to  papers on
 general relativity and cosmology.

\noindent

Among the papers in general relativity, we also will not be able
to comment on all the works of some experts like Prof Fiziev on this field, who wrote scores
 of papers on Heun equations. We will give only the earlier papers and leave
the reader to investigate the later ones in the ArXiv.

\noindent

o In general relativity, in a relatively early work, Teukolsky
studied the perturbations of the Kerr metric \cite {Teukolsky}. If
we take $$ \Psi = \exp(-i\omega t) \exp ( im\phi ) S(\theta) R(r),
$$
for the scalar particle we get two equations.
\begin {equation}
\frac {d}{dr} \Big( \Delta \frac {dR}{d\theta} \Big) + \Big( [
(r^2+a^2)^2 \omega^2 -4aMr\omega m +a^2 m^2 ] \Delta^{-1} -A- a^2
\omega^2 \Big) R=0,
\end{equation}
\begin {equation} \frac {1}{ sin {\theta}} \Big( \frac {d} {d
\theta} sin{\theta} \frac {dS} { d\theta} \Big)  + \Big(
a^2\omega^2 cos^2{\theta} - \frac{ m^2}{ sin^2 {\theta}} +A \Big)
S =0.
\end {equation}
Here $A$ is the separation constant, $ \Delta = r^2 -2Mr +a^2 $ .

\noindent Teukolsky just stated these equations \cite{Teukolsky}. Later these
equations were found to be two coupled singly confluent Heun
equations \cite {batic13}.

o Quasi-normal modes of rotational gravitational singularities
were also studied by solving these equations by E.W. Leaver
\cite{slav87}.

\noindent In recent applications in general relativity, Heun type
equations become indispensable when one studies phenomena in
higher dimensions, or in different geometries.   We must note that even the
simplest black hole metric, the Schwarzschild, has solutions
 in the Heun form \cite {fiziev1} \cite {philipp}.

\noindent
Some other references for general relativity applications
are:

o D. Batic, H. Schmid, M. Winklmeier where the Dirac equation in
the Kerr-Newman metric and static perturbations of the
non-extremal Reisner-Nordstrom solution are studied \cite{schmid}.
D. Batic and H. Schmid also studied the Dirac equation for the
Kerr-Newman metric and looked for its propagator \cite{batic}.
They found that the equation satisfied is a form of a general Heun
equation described in Reference \cite{schmid}. In later work
Batic, with collaborators continued studying Heun equations and
their generalizations \cite{batic2}.
In his most recent paper Batic, with collaborators studied \emph{ Semi commuting and commuting operators for the Heun family} \cite {batic3}.

\noindent Prof. P.P. Fiziev studied problems whose solutions are
Heun equations extensively.

o In a paper published in gr-qc/0603003, he studied the exact
solutions of the Regge-Wheeler equation in the Schwarschild black
hole interior \cite {fiziev1}.

o He presented a novel derivation of the Teukolsky-Starobinsky
identities, based on properties of the confluent Heun functions
\cite {fiziev3}. These functions define analytically all exact
solutions to the Teukolsky master equation, as well as to the
Regge-Wheeler and Zerilli ones.

o In a talk given at 29th Spanish Relativity Meeting (ERE 2006),
he depicted in more detail the exact solutions of Regge-Wheeler
equation,which described the axial perturbations of Schwarzschild
metric in linear approximation, in the Schwarzschild black hole
interior and on Kruskal-Szekeres manifold in terms of the
confluent Heun functions \cite{fiziev6}.

o All classes of exact solutions to the Teukolsky master equation
were described in terms of confluent Heun functions in Reference
\cite {fiziev5} \cite{Denitza}.

o In reference \cite {fiziev4} he reveals important properties of
the confluent Heun's functions by deriving a set of novel
relations for confluent Heun's functions and their derivatives of
arbitrary order. Specific new sub classes of confluent Heun's
functions are introduced and studied. A new alternative derivation
of confluent Heun's polynomials is presented.

o In another paper \cite{fiziev7} he, with a collaborator, noted
that weak gravitational, electromagnetic, neutrino and scalar
fields, considered as perturbations on Kerr background satisfied
Teukolsky Master Equation. The two non-trivial equations were
obtained after separating the variables, one equation only with
the polar angle and another using only the radial variable. These
were solved by transforming each one into the form of a confluent
Heun equation.

o Fiziev is an expert in this topic. Two further articles by him
and his collaborator are \emph{ Solving systems of transcendental
equations involving the Heun functions}, \cite{fiziev8} and
\emph{Application of the confluent Heun functions for finding the
quasinormal modes of non rotating black holes} \cite{fiziev9}.

We also cite one of the last papers of Fiziev on the mathematical
properties of this  subject which can have applications in
physics. In \cite{fiziev111}, the author "introduces and studies a
novel type of solutions to the general Heun equation". His approach
is based " on the symmetric form of the Heun differential equation
yielded by development of the Papperitz-Klein symmetric form of
the Fuchsian equations with an arbitrary number  of regular
singular points greater than 4. The symmetry group of these
equations turns to be a proper extension of the Mobius group". He
also introduces and studies" new series solutions   and
derives solutions for the four singular point case which treats
simultaneously and on an equal footing all singular points."

\noindent Among other papers on this subject one may cite the
following papers:

o R.Manvelyan, H.J.W. Muller Kirsten, J.Q. Liang and Y. Zhang,
calculated the absorption rate of a scalar by a D3 brane in ten
dimensions in terms of modified Mathieu functions, and obtained
the S-matrix in reference \cite{manvelyan}.

o T.Oota and Y.Yasui studied the scalar laplacian on a wide class
of five dimensional toric Sasaki-Einstein manifolds, ending in two
Heun's differential equations in reference \cite{oota}.

o S.Musiri and G. Siopsis found out that the wave equation,
obtained in calculating the asymptotic form of the quasi-normal
frequencies for large AdS black holes in five dimensions, reduces
to a Heun equation, in reference \cite{Musiri}.

o A. Al-Badawi and I. Sakalli studied the Dirac equation in the
rotating Bertotti-Robinson spacetime \cite{sakalli} ending up with
a Heun type equation.

\noindent I first \emph{encountered} this type of equation when we
tried to solve the scalar wave in the background of the Nutku
helicoid instanton \cite{Yavuz2}. In this case, for a scalar
particle in this background metric, one gets the Mathieu equation
which is a special case of the Heun equation. In the same paper,
the solutions in four dimensions involve the product of two
exponentials and two Heun functions. These solutions can be summed
to give the Green's function for  this problem in a closed form.
We could not succeed to obtain a closed form solution for the
Greens function when the similar problem is studied in five
dimensions \cite {tolga1, tolga3}.

o The helicoid instanton is a double-centered solution. As
remarked above, for the simpler instanton solution of
Eguchi-Hanson \cite{ Eguchi} hypergeometric solutions are
sufficient \cite{Nuri}. Here one must remark that another paper
using the Eguchi-Hanson metric ends up with the confluent Heun
equation \cite {Malmendier}. These two papers show that sometimes
judicious choice of the coordinate system and separation ansatz
matters.

o Sucu and \"{U}nal also obtained closed solutions for the spinor
particle written in the background of the Nutku helicoid instanton
\cite{Nuri}, whereas using the separation of variables method
gives us an infinite series of product of two Mathieu functions
\cite {tolga1}

o One can show that the solutions of Sucu and \"{U}nal can be
expanded in terms of Mathieu functions if one attempts to use the
separation of variables method. as described by L.Chaos-Cador and
E. Ley-Koo \cite{Chaos}.

o  Tolga Birkandan  and I also found an extension of the Heun
equation with five singular points \cite{tolga2}, and calculated
the solution of a scalar field in the background of the
Eguchi-Hanson equation trivially extended to five dimensions
\cite{tolga2}. Then the solution for  the radial component turned
out to be given in terms of the confluent Heun equation.

o Mirjam Cveti\v{c} and Finn Larsen studied grey body factors and
event horizons for rotating black holes with two rotation
parameters and five charges in five dimensions. When the Klein-Gordon equation for a scalar particle in this background is
written, one gets a confluent Heun equation. In the asymptotic
region this equation turns into the hypergeometric form
\cite{mirjam1}. When they studied the similar problem for the
rotating black hole with four $U(1)$ charges, they again obtained
a confluent Heun equation for the radial component of the Klein-Gordon equation, which they reduce to the hypergeometric form by
making approximations \cite {mirjam2}. These two papers are
partly repeated in \cite {mirjam3} . Same equations were obtained
which were reduced to approximate forms which gave solutions in
the hypergeometric form.

\noindent M. Cveti\v{c} encounters this function in several of her
publications and reduces them to the hypergeometric form by giving
physical arguments to drop certain terms in the equation. The
hypergeometric solution points to the presence of conformal
symmetry in the reduced model \cite{mirjam4,mirjam5}. The method
is going to the extreme and near extreme (Kerr/CFT correspondence)
limits, going to the boundary and in some cases using a
``subtracted metric" using a warp factor which preserves all the
near horizon properties of the black hole such as the entropy and
the thermodynamic potentials, and if necessary dropping certain
terms which are negligible in these limits
\cite{mirjam6,mirjam7,mirjam8}.

\noindent \emph{``In general, conformal symmetry does not exist in
the non-extremal cases. The solutions often turn out to be of the
Heun form. In the extremal case two horizons overlap. In the near
extremal case they are very close to each other. In these two
cases and in the near horizon limit, we find conformal symmetry,
resulting in solutions which are hypergeometric functions, or one
of its confluent forms. If we want conformal symmetry without
going to the extremal or the near horizon limit, we have to change
the  `warp factor'.  When the warp factor is changed, the rest of
the metric preserves its initial form. The thermodynamic
potentials and entropy do not change. You have to drop some terms
resulting in solutions in the hypergeometric form. This is
equivalent to putting the black hole into a conic box.  If you go
to the asymptotic or to the scaling limit, this is seen clearly.
In these limits the Einstein equations are not satisfied unless
the energy-momentum tensor, on the right side of the Einstein
equations are also changed, to account for putting the system into
the conic box". \cite{Birkandan2} }

\noindent Cveti\v{c} also studied black holes in supergravity with
Birkandan. Heun solutions also exist for the Wu Black Hole which
is the most general solution of maximally supersymmetric gauged
supergravity in D=5 \cite {Birkandan}. Here they did not study the
limiting cases.
 For the
massless Klein-Gordon equation in the background of the most
general black hole in four dimensions and N=2 gauge supersymmetry
with $U(1)^2$ gauge symmetry  (Chow-Compere solution \cite
{Chow}), the angular equation gives Heun type solutions. The
radial equation has five regular singularities, which reduce to
hypergeometric functions in the near horizon extremal limit \cite
{Birkandan1}.

o We should also mention two papers by H.R. Christiansen and M.S.
Cunha with Heun type solutions. These are: \emph{Confluent Heun
functions in gauge theories on thick braneworlds} \cite{Cunha1},
and  \emph{Kalb-Ramond excitations in a thick-brane scnario with
dilaton} \cite{Cunha2}. In the first paper, the propagation modes
of gauge fields in an infinite Randall-Sundrum scenario are
investigated. Here a sine-Gordon soliton represents the thick four
dimensional braneworld while an exponentially coupled scalar field
acts for the dilaton. For the gauge field motion a differential
equation is found which can be transformed into a confluent Heun
equation.  In the second paper a similar scenario is used. Here a
bulk Kalb- Ramond field is coupled to a dilaton, in a warped
space-time in the presence of a brane field in five dimensions.
Full spectrum and eigenstates are studied.  In the general case,
the solution to the field equations are given in terms of the
confluent Heun function, which reduces to the confluent
hypergeometric function for special values of the parameters.

\noindent Other relevant references I could find, are listed as
references \cite {siopsis} - \cite {Yoshida}.

\noindent o  The more recent papers on this subject include \emph{
The quantum treatment of the 5D-warped Friedman-Robertson-Walker
universe in Schrodinger Picture }\cite {Dariescu}. Here the
time-evolving Schrodinger version of the Wheeler-De Witt equation,
written for the five dimensional warped k=0-FRW Universe is
studied. For small values of the cosmological scale factor, $a$,
the wave function of the Universe is expressed in terms of the
Heun Double Confluent functions, whereas for large  values of this
parameter the solution becomes the Hermite associated functions.
Two papers by the same authors using Heun type functions are
\emph{Fermions in magnestar's crust in terms of Heun double
confluent functions} \cite{Dariescu1}, and \emph{The approximative
analytic study of fermions in magnetar's crust; ultra-relativistic
plane waves, Heun and Mathieu solutions and beyond }
\cite{Dariescu2}.

\noindent o In \emph{Fermi surfaces and analytic Green's functions from
conformal gravity} \cite {Hai}, T2-symmetric charged AdS black
holes are constructed in conformal gravity. The most general
solution up to an overall conformal factor contains three
non-trivial parameters: the mass, electric charge and a quantity
that can be identified as the massive spin-2 hair. The Dirac
equation for the charged massless spinor in this background  can
be solved in terms of the general Heun's function for generic
frequency $\omega$ and wave number $k$. This allows us to obtain
the analytic Green's function $G(\omega, k)$ for both extremal and
non-extremal black holes. For some special choice of black hole
parameters, the Green's function reduces to simpler hypergeometric
or confluent hypergeometric functions.

\noindent o Two of the authors of the paper quoted above had calculated the
Greens's functions in terms of the Heun function in an earlier
paper,  \emph{Exact Green's functions from conformal gravity}
\cite{Lu}.

\noindent o Another paper is:\emph{ Quantized black hole and Heun
function} by D. Momeni, K. Yerzhanov and R. Myrzakulov
\cite{momeni} where a black hole is quantized using the Bohr
method. The solution turns to be of the Heun type .

\noindent o In \emph{On an approach to constructing static ball
models in general relativity} by A.M. Baranov, some solutions of
the Einstein equation were described by Heun functions
\cite{Baranov}.

\noindent o In an unpublished paper, \emph{On analytic solutions of wave equations in regular coordinate systems on
Schwarzschild background}
Dennis Philipp and Volker Perlick claim that they find   " The wave equation for the propagation of (massless) scalar, electromagnetic and gravitational waves on fixed
Schwarzschild background spacetime, which  is described by the general time-dependent Regge-Wheeler equation,
canbe transformed  to usual Schwarzschild, Eddington-Finkelstein, Painleve  Gullstrand
and Kruskal-Szekeres coordinates. In the first three cases, but not in the last one, it is
possible to separate a harmonic time-dependence. Then the resulting radial equations belong to the
class of confluent Heun equations" \cite{Perlick}.

\noindent Among additional papers we can also cite the article of
Bezerra et al, \emph{Exact solutions of the Klein-Gordon equation
in the Kerr-Newman background and Hawking Radiation} where both
the radial and angular solutions are given in terms of confluent
Heun functions \cite{Vieira}. In the particular case corresponding
to an extreme Kerr-Newman black hole, the solution is given by the
double confluent Heun functions \cite{Vieira1}. Biconfluent Heun
functions were obtained for the exact solution of the Schrodinger
equation for a particle (galaxy) moving in a Newtonian universe
with a cosmological constant \cite{Vieira2}.

\noindent Other papers on general relativity written in 2015 also
include \emph{New results for electromagnetic quasinormal and
quasibound modes of Kerr black holes}, by D.Staicova and P.Fiziev
\cite{Staicova},  where the authors solve Teukolsky equations with
confluent Heun solutions numerically. In \emph{Heun functions
describing fermions evolving in paralel and magnetic fields}, by
C. Dariescu and M.A. Dariescu, \cite {M.A.},  the solutions
are in terms of double confluent Heun functions. Same authors also
published \emph{ Quantum analysis of k=-1 Robert-Walker Universe},
where they solved the Wheeler-DeWitt equation \cite{M.A.1}. The
solutions turned out to be Heun functions.  M.C. E. Cedeno and
 J.C.N. Araujo show that for Master equation solutions in the linear regime of characteristic
formulation of general relativity, the  solution is in terms of
confluent Heun's functions for radiative  case in the
Schwarzschild's background \cite{Cedeno}. In \emph{Massless Dirac
particles in the vacuum C-metric}, D.Bini et al show that the Dirac
equation, written in the background of the C- metric can be
reduced to a radial and an angular equation, both of which can be
solved in terms of general Heun functions \cite{Bini}. Vieira et
al\cite{Silva} show that for \emph{Charged massive scalar fields are
considered in the gravitational and electromagnetic field produced
by a dyonic black hole with a cosmic string along its axis of
symmetry} " exact solutions of both angular and radial parts of the
covariant Klein-Gordon equation in this background can be obtained,
and are given in terms of the confluent Heun functions". In
\cite{Kofron1} , Kofron separates test fields equations on the
non-rotating C- metric background. He finds that the resulting
equations are of the Heun or confluent Heun form for the general
case. These equations, however, can be reduced to hypergeometric
functions in the static, axial symmetric and the extremal case
where the inner and outer horizons coalesce. In another paper
\cite{Kofron2}, the same author studies the similar phenomena on
the background of the rotating C- metric. For the general case,
the radial equation has five regular singularities. In the
extremal, static and axial symmetric cases, one obtains a
polynomial solution.

Some other papers  published in 2016 in the field of general
relativity where solutions  to field equations in the background
of different metrics are as follows:

\noindent
 Valtancoli \cite{Valtancoli} found Heun
solutions for the radial part of the Klein-Gordon equation when
 the scalar field is conformally coupled to a charged BTZ black hole.

\noindent
Vieira and Bezerra \cite{Vieira3} study  "resonant frequencies,
Hawking radiation and scattering of scalar waves...", and find
confluent Heun solutions. They also study \cite{Vieira4} the class
of solutions of the Wheeler-DeWitt equation in the
Friedmann-Robertson-Walker universe. In still another paper
\cite{Vieira5}, these authors find confluent Heun solutions for
the massless Klein-Gordon equation  in the background metric of
the three dimensional rotating and four dimensional canonical
acoustic black holes.

\noindent
 Sakalli \cite{sakalli25} finds analytical solutions in rotating
linear dilaton black holes.

\noindent
Kraniotis \cite {G.V.} studies the
Klein-Gordon equation in the background metric of the Kerr-Newman
(anti) de Sitter black hole. He first reduces the radial and
angular equations to the Heun form, writes the solution in terms
of local Heun   and confluent Heun functions. In my opinion this
paper should be also praised for the introduction of the "false
singular point" concept, which reduces the solution to
hypergeometric functions for certain values of the physical
parameters in the equation.

Since we updated this paper in February 2017, we find close to  thirty new publications if one searches the word “ Heun Functions” in the index  Web of Knowledge ln April 2018. Many of these papers are on the mathematical aspects of the equation and solving Schrodinger equations for different new potentials in terms of Heun or linear combinations of Heun functions. There are also solutions in terms of Heun functions for equations used in different branches of physics. Here we will attempt to review only the papers for applications in physics related to general relativity.

In \cite{Tezcan}, Arda et al  solve the energy relations obtained with the help of the quantization rule for the Klein-Gordon equation with a linear plus an inverse-linear potential in terms of bi-confluent Heun equations.
Vieira wrote two papers \cite {H.S.1, H.S.2} where he first studied \emph{Resonant frequencies of a hydrodynamic vortex}. The radial equation has solutions in terms of double confluent Heun functions. In the second paper, analytic solutions  for sound perturbations in the presence of a rotating acoustic black hole which is an analogue of the conical Kerr metric  were studied. In the massless case, the radial equation has Heun type solutions. 
Vieira also wrote another paper with co-authors \cite {H.S.3}, where \emph{massive scalar fields are considered in the gravitational field produced by a Schwarzschild black hole with a global monopole in f(R) gravity}. The exact solution of the radial part of the Klein-Gordon equation in this background is given in terms of the general Heun functions. The properties of the general Heun functions are applied to study the Hawking radiation and the resonant frequencies of scalar particles.

Ciprian Dariescu wrote two papers with collaborators \cite{Ciprian1, Ciprian2}. In the the first paper, using a perturbative method, Klein-Gordon equation for a charged massive field in the background of a magnetar is solved both in the interior solution and outside the star. Equations can be seperated with general and confluent Heun function solutions. With special conditions on parameters, polynomial solutions can be found and first order transition amplitudes are computed \cite
{Ciprian1}.
In the second paper, for the spatially open Friedmann-Robertson-Walker (FRW) Universe with stiff matter and radiation as non-interacting matter sources, the scale function coming from the integration of the Friedmann equation is expressed in terms of elliptic integrals. For a negative cosmological constant, the allowed ranges for the model’s parameters are identified. Within the quantum analysis, the Wheeler– DeWitt (WDW) equation turns into a modified Morse equation whose solutions are Mathieu and Heun functions. 
\cite{Ciprian2}. 

Sobhani et al \cite{Sobhani} wrote a paper where the thermodynamicl properties of the anharmonic oscillator cosmic string framework are studied. The Schrodinger equation is written in the cosmic string framework and anharmonic oscillations are investigated. The wave function and energy spectrum are derived using confluent Heun functions.
 
Birkandan was also active in this period. He wrote four papers. In the  first paper, with Bouaziz, he showed that the deformed Schrodinger equation for a singular inverse square potential in coordinate space with a minimal length is solved in terms of Heun functions \cite{Tolgab1}.
In his second paper with a collaborator, confluent Heun solutions to the radial equations of two Halilsoy-Badawi metrics are found. For the first metric, the radial part of the massless Dirac equation and for the second case,  the radial part of the massless Klein-Gordon equation  are studied \cite{Tolgab2}, both with Heun type solutions.
In the third paper, he and his collaborator showed that Heun-type exact solutions emerged for both the radial and the angular equations for the case of a scalar particle coupled to the zero mass limit of both the Kerr and Kerr-(anti)de-Sitter space times. Since any type D metric has Heun-type solutions, it is interesting that this property is retained  when the black hole has a zero mass limit. This work further refuted the claims that mass of the black hole, going to zero limit of the Kerr metric was both locally and globally the same as the Minkowski metric \cite{Tolgab3}.
We comment on the fourth paper in the Conclusion section.

\noindent A comprehensive bibliography can be found at the
bibliography section of http://tcpa.uni-sofia.bg/heun/home.html,
compiled by Profs. Plamen Fiziev and Denitsa Staicova.

\noindent Just to give an example of how the Heun function is
emerges in a simple problem, in the next section, our work in
\cite{tolga2}  for the scalar particle in the background metric of
the extended Eguchi-Hanson solution will be sketched..

\section{Scalar field in the background of the extended
Eguchi-Hanson solution}

\noindent\ To go to five dimensions, we can add a time component
to the Eguchi-Hanson \cite{Eguchi} metric so that we have
\begin{equation}
ds^{2}=-dt^{2}+{\frac{{1}}{{1-{\frac{{a^{4}}}{{r^{4}}}}}}}%
dr^{2}+r^{2}(\sigma _{x}^{2}+\sigma _{y}^{2})+r^{2}(1-{\frac{{a^{4}}}{{r^{4}}%
}})\sigma _{z}^{2}
\end{equation}%

\noindent where
\begin{equation}
\sigma _{x}={\frac{{1}}{{2}}}(-\cos \xi d\theta -\sin \theta \sin
\xi d\phi )
\end{equation}%
\begin{equation}
\sigma _{y}={\frac{{1}}{{2}}}(\sin \xi d\theta -\sin \theta \cos
\xi d\phi )
\end{equation}%
\begin{equation}
\sigma _{z}={\frac{{1}}{{2}}}(-d\xi -\cos \theta d\phi ).~
\end{equation}%

\noindent This is a vacuum solution. If we take
\begin{equation}
\Phi =e^{ikt}e^{in\phi }e^{i(m+{\frac{{1}}{{2}}})\xi }\varphi
(r,\theta ),
\end{equation}%

\noindent we find the scalar equation as
\begin{eqnarray}
\varphi (r,\theta ) &=&({\frac{{r^{4}-a^{4}}}{{r^{2}}}}\partial _{rr}+{%
\frac{{3r^{4}+a^{4}}}{{r^{3}}}}\partial _{r}+k^{2}r^{2}+{\frac{{4a^{4}m^{2}}%
}{{a^{4}-r^{4}}}}+ \notag \\
&&4\partial _{\theta \theta }+4\cot \theta \partial _{\theta }+{\frac{{%
8mn\cos \theta -4(m^{2}+n^{2})}}{{\sin ^{2}\theta }})}\varphi
(r,\theta ).
\end{eqnarray}

%\begin{eqnarray}
%\varphi (r,\theta ) =\Big({\frac{{r^{4}-a^{4}}}{{r^{2}}}}\partial _{rr}+{%
%\frac{{3r^{4}+a^{4}}}{{r^{3}}}}\partial _{r}+k^{2}r^{2}+{\frac{{4a^{4}m^{2}}%
%}{{a^{4}-r^{4}}}}+ \notag \\
%&&4\partial _{\theta \theta }+4\cot \theta \partial _{\theta }+{\frac{{%
%8mn\cos \theta -4(m^{2}+n^{2})}}{{\sin ^{2}\theta }}Big))\varphi
%(r,\theta ).
%\end{eqnarray}

\noindent If we take $\varphi (r,\theta )=f(r)g(\theta )$, the
solution of the radial part is expressed in terms of confluent
Heun ($\mathit{H}_{C}$) functions.
\begin{equation*}
f\left( r\right) =\left( -a^{4}+r^{4}\right)
^{{\frac{{1}}{{2}}}\,m}\
\mathit{H}_{C}\left( 0,m,m,{\frac{{1}}{{2}}}\,{k}^{2}{a}^{2},{\frac{{1}}{{2}}%
}\,{m}^{2}-{\frac{{1}}{{4}}}\,\lambda -{\frac{{1}}{{4}}}\,{k}^{2}{a}^{2},\,{{%
\frac{{{a}^{2}+{r}^{2}}}{{2{a}^{2}}}}}\right) \
\end{equation*}%
\begin{equation}
+\left( {a}^{2}+{r}^{2}\right) ^{-{\frac{{1}}{{2}}}\,m}\left(
r^{2}-a^{2}\right) ^{{\frac{{1}}{{2}}}\,m}\mathit{H}_{C}\left( 0,-m,m,{\frac{%
{1}}{{2}}}\,{k}^{2}{a}^{2},{\frac{{1}}{{2}}}\,{m}^{2}-{\frac{{1}}{{4}}}%
\,\lambda -{\frac{{1}}{{4}}}\,{k}^{2}{a}^{2},{{\frac{{{a}^{2}+{r}^{2}}}{{2{a}%
^{2}}}}}\right)
\end{equation}%

\noindent If the variable transformation $r=a\sqrt{\cosh x}$ is
made, one solution can be expressed as
\begin{equation}
f\left( x\right)= { \left( \sinh \left( x\right)
\right)^{m}\ \mathit{H}_{C}\left( 0,m,m,{\frac{{1}}{{2}}}\,{k}^{2}{a}%
^{2},{\frac{{1}}{{2}}}\,{m}^{2}-{\frac{{1}}{{4}}}\,\lambda -{\frac{{1}}{{4}}}%
\,{k}^{2}{a}^{2},{\frac{{1}}{{2}}}\,\cosh^2 (x/2)\right).}
\end{equation}

\noindent We tried to express the equation for the radial part in terms of $%
u={\frac{{a^{2}+r^{2}}}{{2a^{2}}}}$ to see the singularity
structure more clearly. Then the radial differential operator
reads
\begin{equation}
4{\frac{{d^{2}}}{{du^{2}}}}+4\left( {\frac{{1}}{{u-1}}}+{\frac{{1}}{{u}}}%
\right) {\frac{{d}}{{du}}}+k^{2}a^{2}\left( {\frac{{1}}{{u-1}}}+{\frac{{1}}{{%
u}}}\right) +{\frac{{m^{2}}}{{u^{2}(1-u)^{2}}}}.
\end{equation}

\noindent This operator has two regular singularities at zero and
one, and an irregular singularity at infinity, the singularity
structure of the confluent Heun equation. This is different from
the hypergeometric equation, which has regular singularities at
zero, one and infinity.

\qquad \noindent The  solution of the angular  equation which is
regular at $\theta = \pi$  for $m$ greater than $n$ is given below
in terms of hypergeometric functions.
\begin{equation*}
g(\theta) = \sin(\theta)^{m} \cot(\theta/2)^{n}
\end{equation*}%
\begin{equation*}
\times {\mathit{_{2}F_{1}}(([m+{\frac{{1}}{{2}}}\,\sqrt{\lambda +1}+{\frac{{1}%
}{{2}}},m-{\frac{{1}}{{2}}}\,\sqrt{\lambda +1}+{\frac{{1}}{{2}}}],)\ {%
[1+n+m],{\frac{{1}}{{2}}}\,\cos^2 \left( \theta \right)/2 })}.
\end{equation*}

\section {Conclusion}

\noindent In this paper, first the  Heun function is introduced, then some its uses in physics, especially in the field
 of general relativity and gravitation are demonstrated.  We have to note that most of the physicists that bluntly state their solution
 is in terms of Heun functions  are mainly from the third world. We see physicists from Bulgaria, Romania, Brazil, Armenia, even Turkey in this group. There are mathematicians from the western world, though, who are experts in this field. 
  Batic, a mathematician, although he now works in U.A.E. may be considered from  the western world.   Ronveaux from Belgium, and many other mathematicians are from the western world.

They are not really many exceptions to this observation. Cveti\v{c} and
Larsen demonstrate what the physicists from the western world do.
They try to express their solutions in terms of hypergeometric
functions, by going to the asymptotics, to the extremal or to the
near extremal limit, or putting the solution into a conic box, by
changing the energy momentum term if necessary, but keeping the
thermodynamic potentials same. A long endeavor was necessary to
label  \emph{Teukolsky Master Equations} as belonging to the Heun
class \cite{batic13}. Only recently the equation given by 't Hooft
\cite{Hooft} was shown to belong to the Heun class if it were not
modified \cite{Tolga23}.  When modified the solution is the
manageable
 hypergeometric function.
We agree that this  impression may be wrong, but it is just an
observation.

\noindent The first version of this paper was submitted to the
13th Regional Conference on Mathematical Physics, which was held
in Antalya, Turkey on 27-31 October 2010 and printed in
\cite{mahmut}.

\section{ Acknowledgement}

\noindent I am grateful to Prof.s Cemsinan Deliduman and Kayhan \"{Ulker} for providing me a shelter at Mimar Sinan Fine Arts University during my days in retirement.
  I indebted to Tolga Birkandan for collaboration and technical assistance.  I am  grateful to  Prof. Dr.
Andr\'{e} Ronveaux for informing me of a slight error in my
reference 9.
 I thank
Science Academy,  Turkey for support.

\end{document}